\documentclass{article}

\usepackage{arxiv}

\usepackage[utf8]{inputenc} % allow utf-8 input
\usepackage[T1]{fontenc}    % use 8-bit T1 fonts
\usepackage{hyperref}       % hyperlinks
\usepackage{url}            % simple URL typesetting
\usepackage{booktabs}       % professional-quality tables
\usepackage{amsfonts}       % blackboard math symbols
\usepackage{nicefrac}       % compact symbols for 1/2, etc.
\usepackage{microtype}      % microtypography
\usepackage{lipsum}		% Can be removed after putting your text content
\usepackage{graphicx}
\usepackage{natbib}
\usepackage{doi}
\usepackage{amsmath}

\title{Dynamically altered conductance in an Organic Thin Film Memristive Device}

\author{ Agnieszka I. Pawłowska\\
	Faculty of Physics, Astronomy, and Applied Computer Science,\\ Jagiellonian University,\\
    Łojasiewicza 11, 30-348 Krakow, Poland\\
	\texttt{a.pawlowska@doctoral.uj.edu.pl} \\
	\And
	Paweł D\k{a}bczyński \\
	Faculty of Physics, Astronomy, and Applied Computer Science,\\ Jagiellonian University,\\
    Łojasiewicza 11, 30-348 Krakow, Poland\\
	\texttt{pawel.dabczynski@uj.edu.pl} \\
    \And
	Sebastian Lalik \\
	Faculty of Physics, Astronomy, and Applied Computer Science,\\ Jagiellonian University,\\
    Łojasiewicza 11, 30-348 Krakow, Poland\\
	\texttt{sebastian.lalik@uj.edu.pl} \\
    \And
	Juan Pablo Carbajal \\
	Institute for Energy Technology,\\
    OST Eastern Switzerland University of Applied Sciences,\\
    Oberseestrasse 10, 8640 Rapperswil, Switzerland\\
	\texttt{juanpablo.carbajal@ost.ch} \\
    \And
	Dante R. Chialvo \\
	Center for Complex Systems and Brain Sciences (CEMSC${^3}$) \& Instituto de Ciencias Físicas (ICIFI-Conicet),\\ 
    Escuela de Ciencia y Tecnología, Universidad Nacional de Gral. San Martín,\\ 
    Campus Miguelete, 25 de Mayo y Francia (1650), \\
    San Martín, Buenos Aires, Argentina\\
	\texttt{dchialvo@gmail.com} \\
    \And
	Jakub Rysz\\
	Faculty of Physics, Astronomy, and Applied Computer Science,\\ Jagiellonian University,\\
    Łojasiewicza 11, 30-348 Krakow, Poland\\
	\texttt{jakub.rysz@uj.edu.pl} \\
}

\hypersetup{
pdftitle={Dynamically altered conductance in an Organic Thin Film Memristive Device},
pdfsubject={organic memristive device},
pdfauthor={Agnieszka I. Pawłowska, Paweł D\k{a}bczyński, Sebastian Lalik, Juan Pablo Carbajal, Dante R. Chialvo, Jakub Rysz},
pdfkeywords={organic memristive devices, synaptic plasticity, poly(3-hexylthiophene)},
}

\begin{document}
\maketitle

\begin{abstract}
The memristive device is one of the basic elements of novel, brain-inspired, fast, and energy-efficient information processing systems in which there is no separation between memorization and information analysis functions.
Since the first demonstration of the resistive switching effect, several types of memristive devices have been developed.
In most of them, the memristive effect originates from direct modification of the conducting area, e.g. conducting filament formation/disintegration, or semiconductor doping/dedoping. Here, we report a  solution-processed lateral memristive device based on a new conductivity modulation mechanism.
The device architecture resembles that of an organic field-effect transistor in which the top gate electrode is replaced with an additional insulator layer containing mobile ions.
Alteration of the ion distribution under the influence of applied potential changes the electric field, modifying the conductivity of the semiconductor channel.
The devices exhibit highly stable current-voltage hysteresis loops and Short-Term Plasticity (STP).
We also demonstrate short-term synaptic plasticity with tunable time constants.
\end{abstract}

% keywords can be removed

\section{Introduction}
The separation of the memory and processing unit proposed by von Neuman, which lies at the heart of the architecture of modern processors, has contributed to their enormous success, but is now seen as a constraint on the further development of high-speed and energy-efficient information processing systems.
One possibility to overcome the so-called von Neumann bottleneck is the development of data processing solutions inspired by the architecture of the brain, where memorization and information processing are performed by the network of nerve cells~\citep{Jo_2010, Kim_2016}.
Hardware realizations of artificial neural networks are often based on memristive devices, two terminal passive elements that change resistance upon electrical stimuli.
One of its properties is the presence of a hysteresis loop in the current-voltage characteristics linking two features: the memory effect and resistance.
The need for such a component to build a circuit that simulates the nerve cell membrane was pointed out by Hodgkin and Huxley when studying the electrical activity of nerve cells~\citep{Hodgkin_1952}.
Later, the existence of memristor, the fourth passive element that combines the relationship between charge and flux linkage, was postulated by Leon Chua~\citep{Chua_1971}.
Historically, memristive-type current-voltage characteristics have been observed in devices such as arc lamps, coherer, and other devices~\citep{Caravelli_2018}.
In 2008, HP Laboratory reported the physical realization of a memristive device based on resistance switching~\citep{Strukov_2008}.
It should be mentioned that the realization of the Chua memristor concept is still debatable~\citep{Vongehr_2015}.
However, since the first realization mentioned, the topic has gained a lot of interest among researchers not only due to its intriguing features, but also due to its potential implementations.
To this day, the memristive character has been reported in systems utilizing both inorganic and organic materials.
Most of the memristive devices reported so far are based on transition-metal oxides~\citep{Jo_2010, Strukov_2008, Pickett_2012}.
The reproducible formation of conductive filaments is the most common mechanism responsible for the memristive effect~\citep{Jo_2009, Xu_2019, Bessonov_2015}.
Another switching mechanism has been reported for devices consisting of layers of co-sputtered Ag and Si with a gradient in the Ag / Si mixture ratio, the active layer consists of Ag-rich (high conductivity) and Ag-poor (low conductivity) regions where the metal particles are responsible for conducting front formation, which under the applied bias stays in continuous motion and changes the resistance of the whole device~\citep{Jo_2010}.

New memristive systems that use, for example, polymers, organic molecules, perovskites, or nanoparticles have only recently been introduced~\citep{Zeng_2014, Goswami_2017, Xiao_2016, Alibart_2010}.
The processes behind the memristive character of a device may differ depending on attributes, e.g., conductive filament formation~\citep{Xu_2019}, charge trapping / detrapping~\citep{Younis_2013}, reversible redox reactions~\citep{Liu_2016} or ion migration~\citep{Park_2020}.
An interesting example of an organic memristive device based on conducting filament formation has recently been reported in a system utilizing cross-linkable polymer~\citep{Park_2020}.
Neuromorphic arrays based on this system are reported to operate with high electrical and mechanical endurance because of the predefined localized ion migration paths in the polymer medium.
Another example of the developed memristive device is an artificial synapse network based on proton-doped nanogranular SiO$_2$ in the geometry of the laterally coupled transistor.
In this system, memristive effects, measured as post-synapse current flow between the drain and source electrodes, are caused by proton migration under the presynaptic pulse, defined as the pulse applied to the gate ~\citep{Zhu_2014}.
In addition to the mechanisms mentioned above, devices based on effects such as phase transitions~\citep{Younis_2013, Driscoll_2009, Tuma_2016}, and conversion of transport from Schottky emission to direct tunneling~\citep{Ge_2018} have also been developed.
In almost every reported memristive device, there is a reversible switch between two conduction mechanisms, resulting in an on-resistance state and an off-resistance state.
In 2020 Lu et al.~\citep{Lu_2020} show an artificial synapse system based on reduced graphene oxide, where the conductivity is continuously modified during the increase and decrease of voltage.
To the best of our knowledge, to this day it is the only example of such a device reported in the literature.

In 2013, Pickett et al.~\citep{Pickett_2013a} showed that memristive devices could find applications in the construction of neuristors, electronic devices with properties similar to the Hodgkin–Huxley neural model.
Memristive devices for neuromorphic computing must meet requirements regarding scalability, performance, and reliability~\citep{Im_2020}.
Furthermore, such a device should successfully emulate some basic neural functions, including synaptic plasticity, which is responsible for the brain’s ability to learn. Temporal strengthening/depressing of synaptic weights (dubbed  Short-Term Plasticity or STP) which lasts few minutes before it fades away, it is believed to be the basis of information processing in the human brain~\citep{Wang_2017, Li_2013}.
Short-term plasticity has been emulated in artificial devices as  incremental changes in system conductance caused by consecutive pulses of potentiation (Short-Time Potentiation - STP) or depressing pulses (Short-Time Depression - STD)~\citep{Mead_1990, Jo_2010}.

In this work, we present a novel two-terminal lateral memristive device, fabricated by means of deposition from solution, with in-plane conductivity.
The system consists of three layers of polymers, regioregular poly(3-hexylthiophene) (R-P3HT), which can be considered a model polymer semiconductor, insulating poly(4-vinylpyridine) (P4VP) with surface cross-linked with cobalt-based complexes~\citep{Majcher_2018} and poly(sodium 4-styrenesulfonate) (PSS) with or without the addition of potassium trifluoromethanesulfonate (potassium triflate).
Its principle of operation is based on dynamic changes in semiconductor conductivity caused by processes that occur at the interface between PSS and P4VP.
The result is that the change in the resistance of the device, when voltage is applied, is continuous, both when the voltage is increasing and decreasing.
The device exhibits a highly stable current-voltage hysteresis loop.
The dynamics of the device is successfully replicated by a difference-equation model including  the continuous change of resistance and the direction of its changes.
We have also demonstrated the short-term synaptic plasticity of our system with tunable time constants.
This feature, combined with the possibility of using a cost-effective large-scale fabrication process (e.g., roll-to-roll processes) makes the present device an attractive alternative for neuromorphing technology.

\section{Results and discussion}
\begin{figure}
\includegraphics[width=0.9\linewidth]{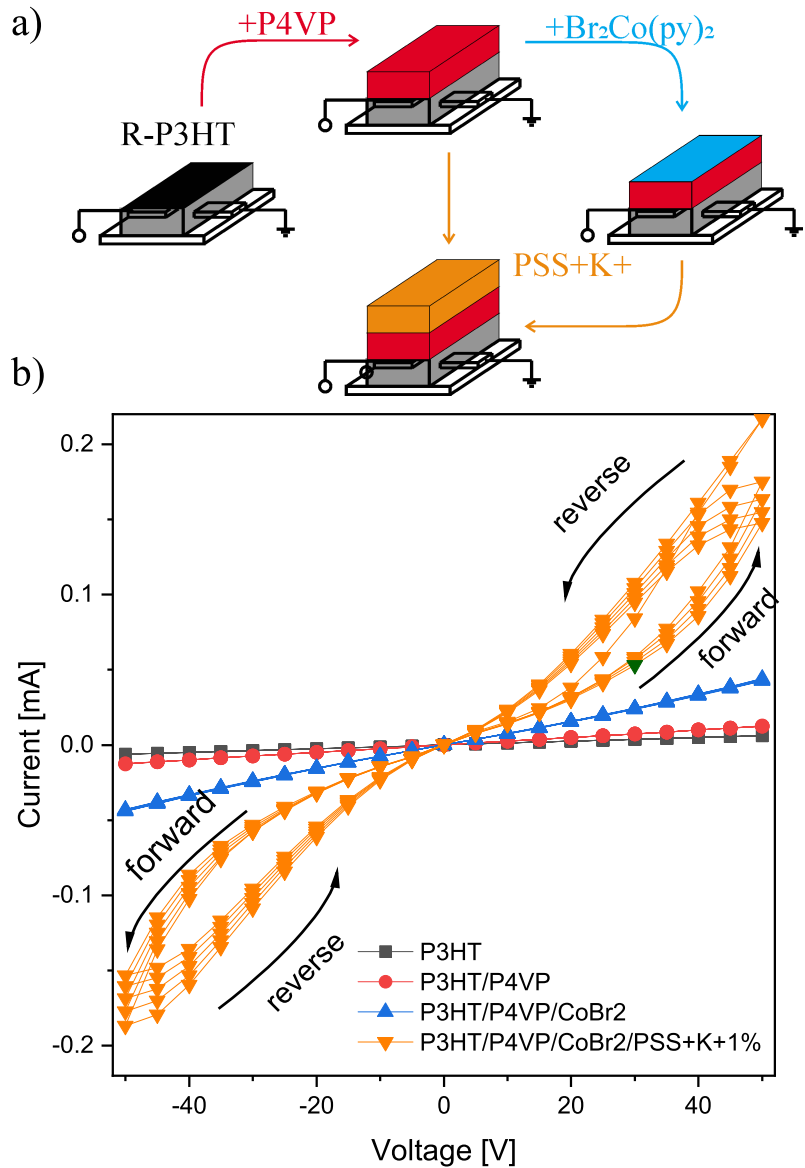}
\caption{\textit{a) A scheme of sample preparation consisting of three (in case of original system) or four (for modified system) separate steps. b) Current-voltage characteristics measured between consecutive preparation steps, the inset shows current-voltage characteristics of a system, measured after each step of preparation, with the R-P3HT layer preparation step omitted }}
\label{fig: fig_1}
\end{figure}

The memristive planar devices were prepared by casting a regioregular poly(3-hexyltiophene) (R-P3HT) solution on a glass substrate with patterned ITO electrodes followed by subsequent deposition of the solution of the poly (4-vinylpyridine) polar insulator (P4VP) and poly( styrene sulfonate) admixed with potassium trifluoromethane sulfonate (potassium triflate) (PSS+1$\%$K$^+$).
In some cases, the surface of P4VP was cross-linked with CoBr$_2$Py$_2$ complexes as described in our previous work~\citep{Majcher_2018, Dabczynski_2020} prior to the deposition of the PSS layer.
\textbf{Figure~\ref{fig: fig_1}a)} illustrates schematically the fabrication steps and structure of two types of devices studied hereafter and referred to as the original structure (without CoBr$_2$ modification) and the Co-modified structure.
All subsequent layers were casted with orthogonal solvents to not cause any damage to the previously casted films (see Supporting Information \textbf{Figure~SI~1)} and special care was taken to preserve conduction in the devices only through the R-P3HT layer.

The thicknesses of the individual layers were the same for all devices tested: $25nm$ R-P3HT,$125nm$ P4VP, $100nm$ PSS(+1$\%$K$^+$), it is worth highlighting that the contact area between the insulating / conducing components was fixed.
\textbf{Figure~\ref{fig: fig_1}b)} illustrates changes in the conductivity of devices after each preparation step, the current-voltage characteristics correspond to: the single layer of R-P3HT (black), R-P3HT / P4VP bilayer (red), R-P3HT / P4VP / CoBr$_2$ (blue) and R-P3HT/P4VP/CoBr$_2$/PSS(+1$\%$K$^+$) (green) multilayers.
Measurements were made in voltage loops in the range of $\pm50V$ with a $2.5V/s$ measurement step starting from $0V$.
As shown in Figure~\ref{fig: fig_1}b), each step of preparation changes the conductivity of the systems; however, the PSS(+1$\%$K$^+$) layer exceptionally changes the character of the conductivity.
The deposition of the P4VP layer increases conductivity due to interactions with the R-P3HT layer.
The further increase in conductivity after CoBr$_2$ crosslinking is attributed to the electric field induced in the semiconductor by the pyridine dipole moments in both the P4VP layer and directed by surface-oriented dipoles associated with Br$_2$CoPy$_2$ complexes~\citep{Dabczynski_2020}.
Presence of the PSS(+1$\%$K$^+$) layer not only increases the conductivity of the system but also incorporates a hysteresis loop into the current-voltage characteristic. The ionic conductivity through the topmost layer was excluded, cf \textbf{Figure SI~2a)}, for devices without an ion reservoir, which is the PSS(+1$\%$K$^+$) layer, only ohmic (linear) conductivity is observed (see \textbf{Figure~SI~2b}).

Dynamic TOF-SIMS profiling experiments (Figure~SI~1) show the absence of potassium ions in the R-P3HT film. This is evidence that the preparation procedure effectively limits the contact of ions from the PSS(+1$\%$K$^+$) reservoir with electrodes.
Additional experiments prove that the ions present in PSS(+1$\%$K$^+$) can migrate in the layer under the applied electrical field.
The results of impedance spectroscopy measurements of a single PSS(+1$\%$K$^+$) layer deposited on the same glass/ITO patterned electrodes reveal relaxation processes both in PSS and PSS(+1$\%$K$^+$) layer.
The addition of K$^+$ ions from the triflate salt modifies the overall time constants of the processes that occur (see supporting information \textbf{Figure~SI~3}).
Furthermore, in the experiment where the surface concentration of K$^+$ was measured with SIMS, where at some point the measurement was paused while a $+50V$ external potential was applied to the sample and then the SIMS measurement was immediately restarted, we can show that after external voltage application, the surface concentration of K$^+$ drops and slowly returns to its initial value (see supporting information \textbf{Figure~SI~4}).
These two findings prove that ions in PSS(+1$\%$K$^+$) have significant mobility and can be moved by external potential.

The data presented above indicate that the current flows in the system only through the R-P3HT layer, whose conductivity varies with the applied potential, and the presence of the layer containing mobile ions is necessary for the hysteresis effect to occur, although these ions do not participate directly in charge transport.
In consideration of these findings, we postulate such a mechanism: i) the permanent dipole moment of pyridine/Br$_2$Co(py)$_2$ complexes as well as ions present in the PSS(+1$\%$K$^+$) layer introduce in the R-P3HT additional electric field, perpendicular to the plane of the layer, this field affects the conductivity of the semiconductor; ii) external potential applied to the electrodes forces current flow in the R-P3HT layer and also modifies the ion distribution in PSS(+1$\%$K$^+$) layer and at the P4VP/PSS(K$^+$) interface (see \textbf{Figure~SI~4}); iii) the mobility of ions is finite so that changes in the component of the electric field, whose source are ions, occur with a certain delay relative to changes in voltage applied to the electrodes; iv) the changes in ionic balance at the interface changes the perpendicular component of the electric field in the R-P3HT layer therefore its conductivity, which leads to the hysteresis loop appearance.

\begin{figure}
\includegraphics[width=\linewidth]{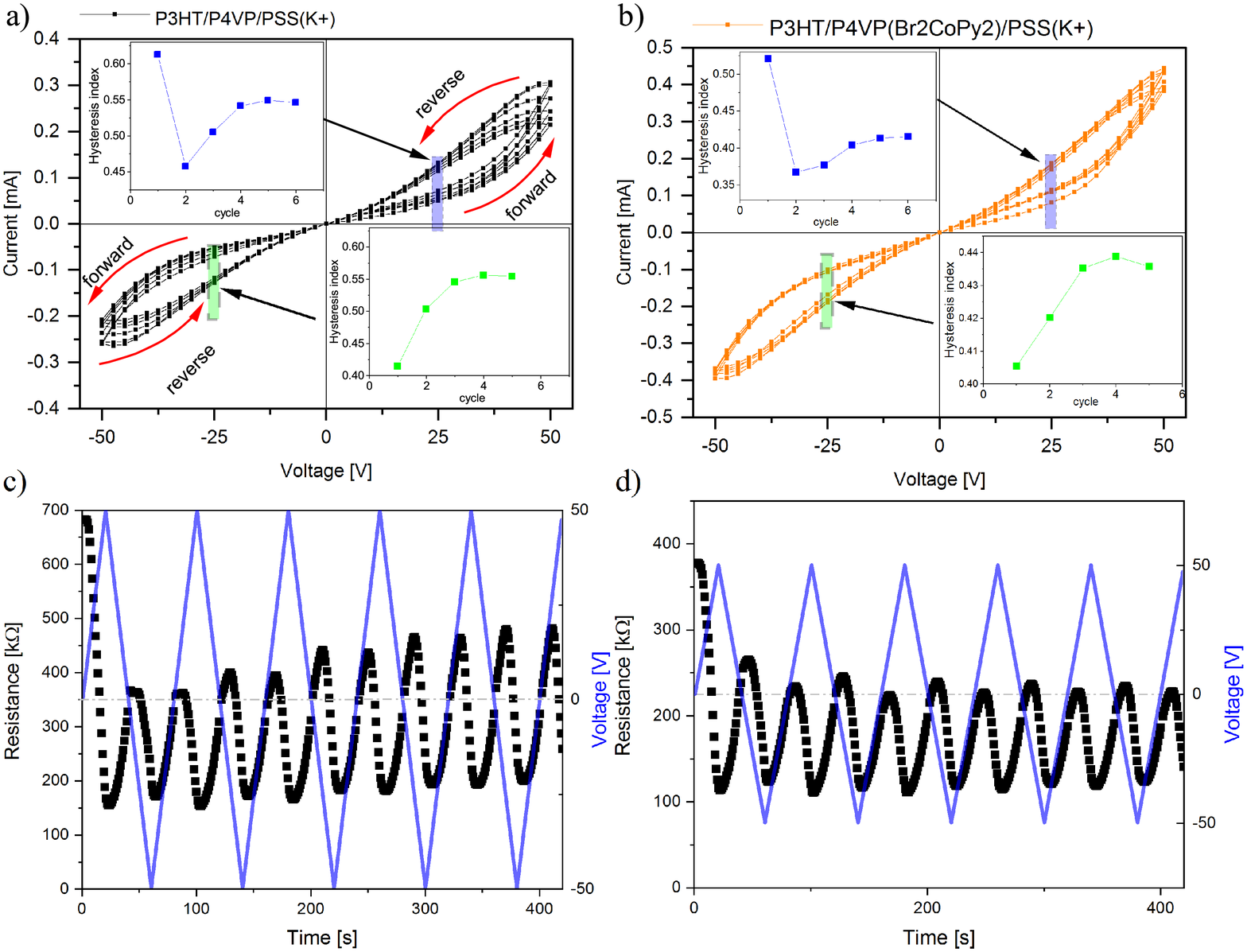}
\caption{\textit{Current–voltage characteristics for a) original system (black) and b) Co-modified system (orange).
Inset plots show Hysteresis Index calculated for $\pm27.5V$.
Calculated resistance of the system as a function of time for c) reference and d) Co-modified system.
The blue line shows the voltage-time curve }}
\label{fig: fig_2}
\end{figure}

\textbf{Figure~\ref{fig: fig_2}} shows the current voltage characteristics for the original (without CoBr$_2$ modification) and Co-modified structures registered for several voltage sweeps.
Currents measured for the Co-modified system are slightly higher for respective voltages, as well as currents stabilize after several measurement cycles, unlike in the original structure, where currents systematically decrease in the consecutive voltage sweeps.
The current-voltage characteristics show that conductivity of both types of device is dynamically altered by applied voltage.

To characterize the loops visible in the first and third quarter of the I-V plane hysteresis index~\citep{Habisreutinger_2018} was calculated:

\begin{equation}
\label{eq:1}
    HI = \left\vert\frac{I_\text{reverse} - I_{forward}}{I_{forward}} \right\vert_{\pm27.5V}.
\end{equation}

The insets in Figures~\ref{fig: fig_2}a) and b) present Hysteresis Index for consecutive measurements (cycles) for negative and positive parts of the I-V curves.
There is an observed dependence of HI on voltage for a given sweep time (see supporting information \textbf{Figure~SI~5}), therefore, in the inset the value for maximum HI obtained for voltage c.a. $\pm27.5V$ is plotted.
The reverse current is defined as the current corresponding to that part of the cycle in which the voltage changes to zero.
Therefore, for forward current, the magnitude of the electric field in the system increases, and for the reverse the magnitude of the electric field decreases in the system.
As can be seen, the hysteresis index in both systems aims at a stable value of c.a. 0.55 for the original system and 0.40 for Co–modified system, respectively.

Despite a slight asymmetry (in the negative and positive part of the I-V curve) in the hysteresis index (c.a. 0.44 for the negative part vs. 0.42 in positive), the behavior of the Co-modified system is similar regardless of the electric field direction.
If we consider the spatial electric field distribution from the two-electrode system of height $h$ (in this case $h$ equals $100nm$), we can show that in the area over the channel between the electrodes the electric field is symmetric with respect to the line perpendicular to the channel center.
On one side, there is a field that attracts the charges to the interface, and on the other side, there is a repulsive force.
Changing the voltage polarity reverses the sides but does not change the overall symmetry picture of the system (a simplified field distribution is shown in the supporting information \textbf{Figure~SI~6}).
This, along with the proposed model, can explain the symmetry in the conduction behavior of the system under positive and negative bias.

Figures~\ref{fig: fig_2} c) and d) show the resistance of the system as a function of the time corresponding to the voltage sweeps as presented in Figure~\ref{fig: fig_2}.
The blue line shows the corresponding time course of the voltage.
For both systems, the resistance oscillates in time and the highest relative change in system conductivity occurs after/during the first voltage sweep.
There is also a phase shift between the resistance of the system and the electric field variation observed for modified and original systems.
Additionally, for the Co-modified system the observed oscillation of the resistance is more stable than for the original.
After 3 cycles, the resistance dependence on the system polarity disappears, as was seen in the HI (see Fig.~\ref{fig: fig_2} a-b insets).
In the original after the 6th cycle still a minor change in the maximum and minimum resistance of the system is observed.
But it should be noted that there is a slight asymmetry between resistance amplitude between $\pm$ polarization.

\begin{figure}
  \includegraphics[width=\linewidth]{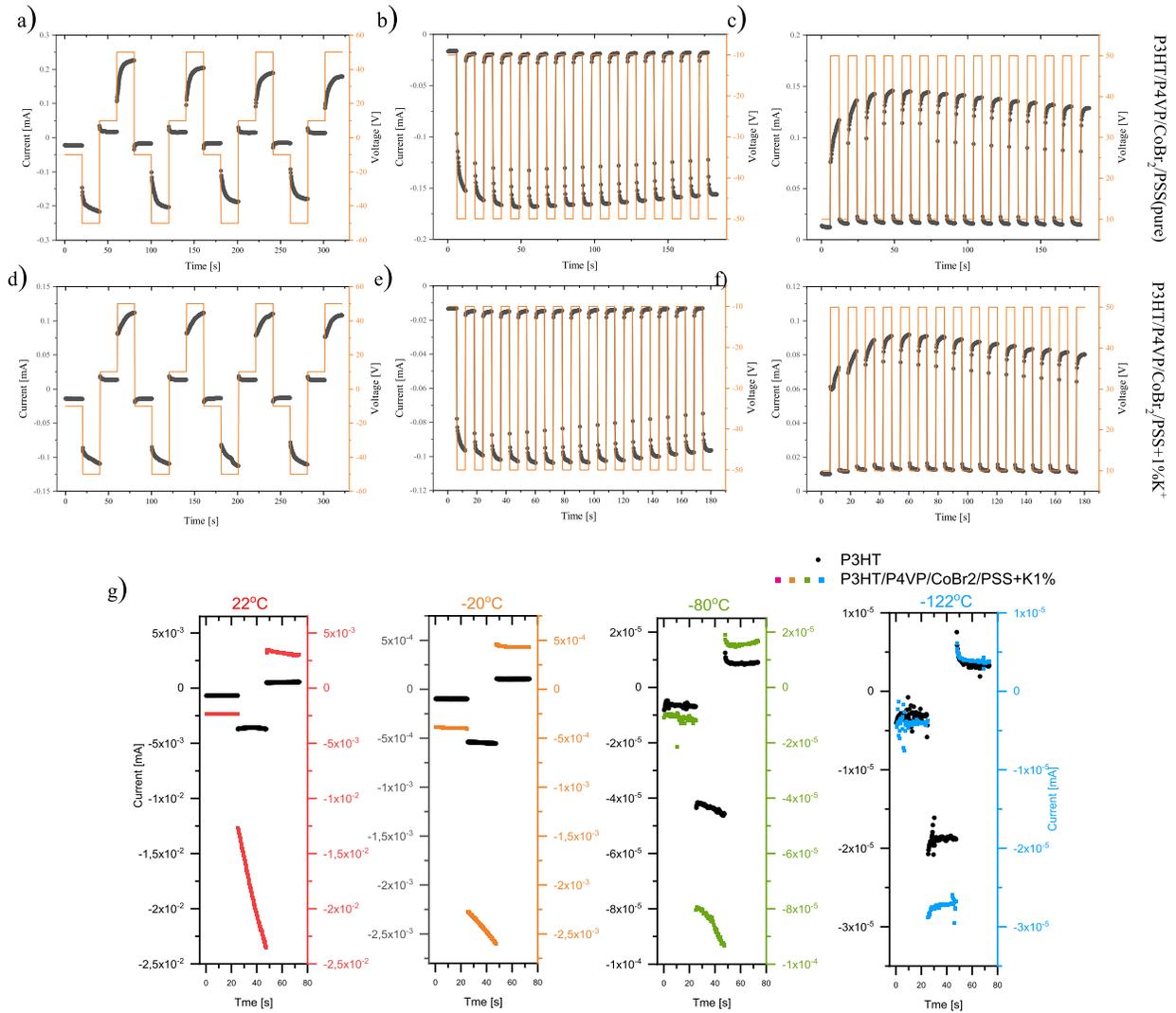}
\caption{\textit{Current vs. time curves registered for three different voltage sequences: a), d) $-10;-50;+10;+50V$, b), e) $-10;-50V$ and c), f) $+10;+50V$ measured for devices with pure PSS matrix (a-c) and devices with matrix enriched with $1\%$ of K$^+$ ions (d-f); the duration of each potential step was $20s$, and the switching time was shorter than $0.25ms$. g) Temperature dependence of devices’ conductivity in 22℃ (red), $-20^{\circ}C$ (orange), $-80^{\circ}C$ (green) and $-122^{\circ}C$ (blue) compared to conductivity of R-P3HT; current response was measured for voltage sequences: $-10;-50;+10V$, the duration of each potential step was $25s$ }}
\label{fig: fig_3}
\end{figure}

To further study the time-constant and neuromorphic characteristics, the Co-modified system was chosen for its improved stability and higher performance.
Two types of devices were tested: systems consisting of pure PSS matrix (\textbf{Figure~\ref{fig: fig_3} a,b,c}) and system with PSS matrix admixed with 1$\%$ of K$^+$ ions (\textbf{Figure~\ref{fig: fig_3}d,e,f}). 
First, the current response to a square wave voltage was measured.
Figure~\ref{fig: fig_3} presents current vs. time curves registered for three different voltage sequences: $-10;-50;+10;+50V$ (Figure~\ref{fig: fig_3}a,d), $+10;+50V$ (Figure~\ref{fig: fig_3}c,f), and $-10;-50V$ (Figure~\ref{fig: fig_3}b,e), the duration of each potential step was $20s$ (Figure~\ref{fig: fig_3}a,d) or $6s$ (Figure~\ref{fig: fig_3}b,c,e,f), and the switching time was shorter than $0.25ms$.
The value of $\pm10V$ was chosen as the “read” potential in the experiments on memory/synaptic properties, since it does not change the conductivity of the system while the current response of the device has a good signal-to-noise ratio.

Figure~\ref{fig: fig_3} shows that the conductivity of both systems increases when $\pm50V$ potential is applied and the respective currents approach $0.23mA$ for the PSS device and $0.12mA$ for PSS+1$\%$K$^+$ device and drops down for $\pm10V$ since respective currents are smaller by an order of magnitude.
For example, for the first voltage sequence the highest currents read from Figure~\ref{fig: fig_3}a) (device with pure PSS matrix) are c.a.: $-0.02mA$ ( $-10V$), $-0.22mA$ ($-50V$), $0.04 mA$ ( $+10V$), 0.23 mA ($+50V$).
Moreover, the conductivity of the whole system decreases globally with each measurement cycle, and the respective currents registered for $+50V$ decrease from approximately 0.23 mA in the first cycle down to approx. $0.18mA$ in the 4th cycle.
Similar/corresponding results have been registered for the system enriched with K$^+$ (Figure~\ref{fig: fig_3}d).
It should be noted that the current response of the devices is symmetrical with respect to the direction of the applied voltage.
The same symmetry is visible when a positive-only or negative-only voltage sweep is applied alternating between $10V$ and $50V$ (Figure~\ref{fig: fig_3}c) or $-10V$ and $-50V$ (Figure~\ref{fig: fig_3}b).
In both cases, during the first three cycles, the current response of the system increases exponentially when +(-)$50V$ is applied and decreases exponentially as the voltage drops down to +(-)$10V$ pulses.

\begin{figure}
  \includegraphics[width=\linewidth]{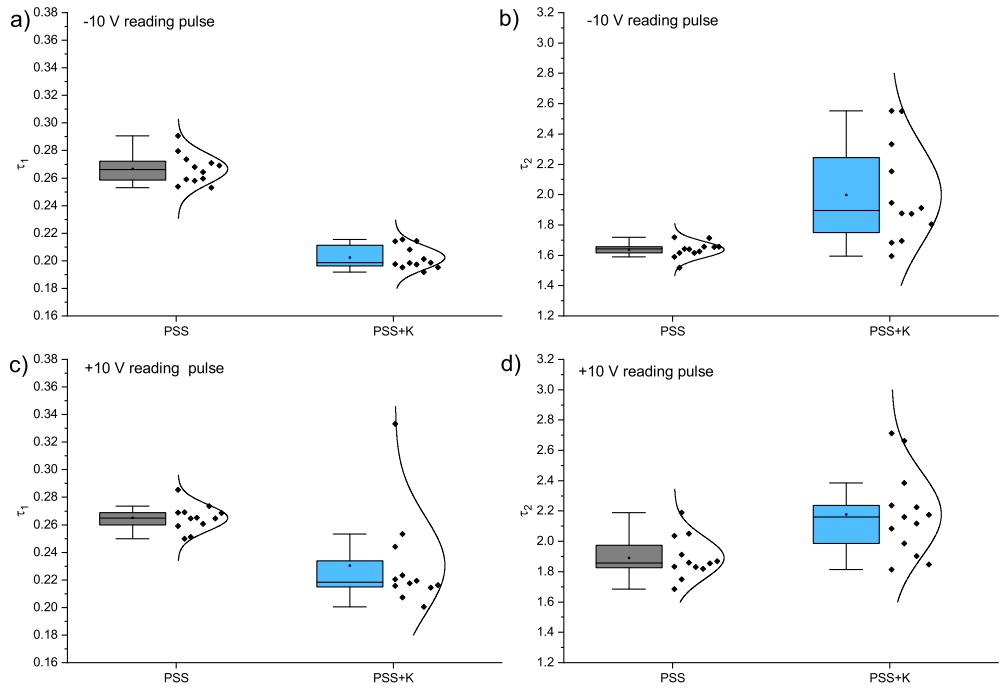}
\caption{\textit{Time constants of relaxation processes calculated by fitting Equation \ref{eq:2} to experimental data for devices consisting of pure PSS and PSS+1$\%$K$^+$ a) $\tau_{1}$ determined for $-10V$ reading pulse, b) $\tau_{2}$ for $-10V$ reading pulse, c) $\tau_{1}$ for $+10V$ reading pulse and d) $\tau_{2}$ for $+10V$ reading pulse}}
\label{fig: fig_4}
\end{figure}

As mentioned above, we believe that changes in conductivity are related to ion diffusion.
Therefore, we have tested the device conductivity at different temperatures in order to determine its changes while the temperature drops, which should result in a decrease in the ion mobility.
Measurements were made for a device consisting of R-P3HT/P4VP/CoBr$_2$/PSS+1$\%$K$^+$ as well as for a sample consisting only of the R-P3HT layer.
\textbf{Figure~\ref{fig: fig_3}g)} shows the results of these measurements carried out at $22^{\circ}C$ (red), -$20^{\circ}C$ (orange), $-80^{\circ}C$ (green), and $-122^{\circ}C$ (blue).
The left-hand side points on the plots are a current response of the semiconductor in corresponding conditions.
As shown on Figure~\ref{fig: fig_3}g conductivity of the whole system differs significantly from the conductivity of the semiconductor itself under different temperature conditions.
In a room temperature, the conductivity of the device increases rapidly with time, while R-P3HT changes its conductivity only slightly.
A similar behavior is observed at $-20^{\circ}C$ and $-80^{\circ}C$, where we observe a decrease in current related to cooling of the semiconductor.
Moreover, we observe that there is some distortion of the current response compared to the corresponding signal in a room temperature.
It might be related to conductivity decrease related to lower current carrier mobility and decreased ion mobility, which, as we believe, is responsible for increased conductivity of the whole system.
At $-122^{\circ}C$ the behavior of the device and of the semiconductor is similar, which indirectly confirms our hypothesis that by lowering ion mobility we can exclude electric field modulation caused by ions and therefore cancel the memristive character of the device.
Quantitatively, the time dependence of the currents registered in Figure~\ref{fig: fig_3}a, d can be described as a double exponential decay corresponding to two coexisting processes of ion relaxation described as follows:

\begin{equation}
\label{eq:2}
    i(t)=A \Big(\exp\Big({\frac{t}{\tau_1}}\Big) + \exp\Big({\frac{t}{\tau_2}} \Big) \Big) + B
\end{equation}
As a result, for a device consisting of a pure PSS matrix, the mean time constants $\tau_{1, +10V} = 0.26s$ and $\tau_{2, +10V} = 1.89s$ for the $+10V$ sequences and $\tau_{1, -10V} = 0.27s$ and $\tau_{2, -10V} = 1.64s$ for the $-10V$ sequences were determined.
For a device consisting of PSS+1$\%$K$^+$ layer average time constants were calculated as $\tau_{1, +10V} = 0.23s$ and $\tau_{2, +10V} = 2.21s$ for the $+10V$ sequences and $\tau_{1, -10V}= 0.20s$ and $\tau_{2, -10V} = 1.99s$ for the $-10V$ sequences.
Total statistical distribution of the time constants for both systems is presented in \textbf{Figure~\ref{fig: fig_4}}.
After the third cycle during $\pm50V$ sequences, the conductivity increases at first and then starts to decrease.
However, globally, the conductivity increases in the first two cycles and decreases with each subsequent cycle.	

\begin{figure}
\includegraphics[width=\linewidth]{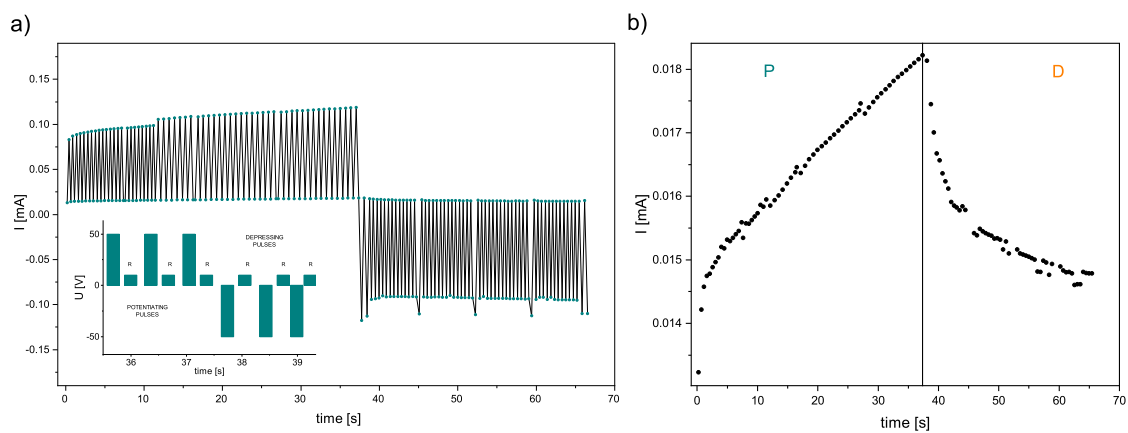}
\caption{\textit{a) Current response to the potentiating and depressing pulses.
The bottom left inset shows the sequence in which the potentiating, reading and depressing pulses were applied; b) the current response to reading pulses of magnitude $+10V$ shown in detail on the right panel with division into responses to potentiating pulses (P) and depressing pulses (D)}}
\label{fig: fig_5}
\end{figure}

Exponential relaxation of electric current is considered one of the key features of a biomimicking memristor for neuromorphic computing; this opens the possibility of modifying its conductivity for a certain time by applying electrical pulses.
Typically, positive pulses increase, while negative pulses decrease the conductivity that corresponds to potentiation or depression of a synapse mimicked by the memristor.
\textbf{Figure~\ref{fig: fig_5}a} shows response of the Co-modified device to series of potentiating (P) ( $+50V$, duration $100ms$) and depressing (D) ( $-50V$ duration $100ms$) pulses, referred as programming pulses~\citep{Jo_2010}.
After each programming pulse, the state of the memristive devices was tested by measuring current under $+10V$ bias, which was measured between the programming pulses, and will be mentioned later in this paper as reading pulses (R) (see inset on Figure~\ref{fig: fig_5}a).
All current measurements were performed with a  $225ms$ time interval between pulses.

As the results show, global conductivity increases/decreases with each pulse during potentiating/depressing cycles regardless of the magnitude of the applied voltage.
It is especially apparent for current responses measured for $+10V$ shown in \textbf{Figure  \ref{fig: fig_5}b}.
These results, as well as those presented in Figure~\ref{fig: fig_3}, indicate that it is possible to program the system response to pulses of lower magnitude with the preceding pulses of higher magnitude. Presented conductivity changes of the devices are temporal due to dynamic processes which occur in the system.
The specific mechanism of operation shown in Figure~\ref{fig: fig_5} has been previously reported in literature as Short-Term Synaptic Plasticity ~\citep{Jo_2010}.

\section{Difference-Equation model}
\begin{figure}
  \includegraphics[width=\linewidth]{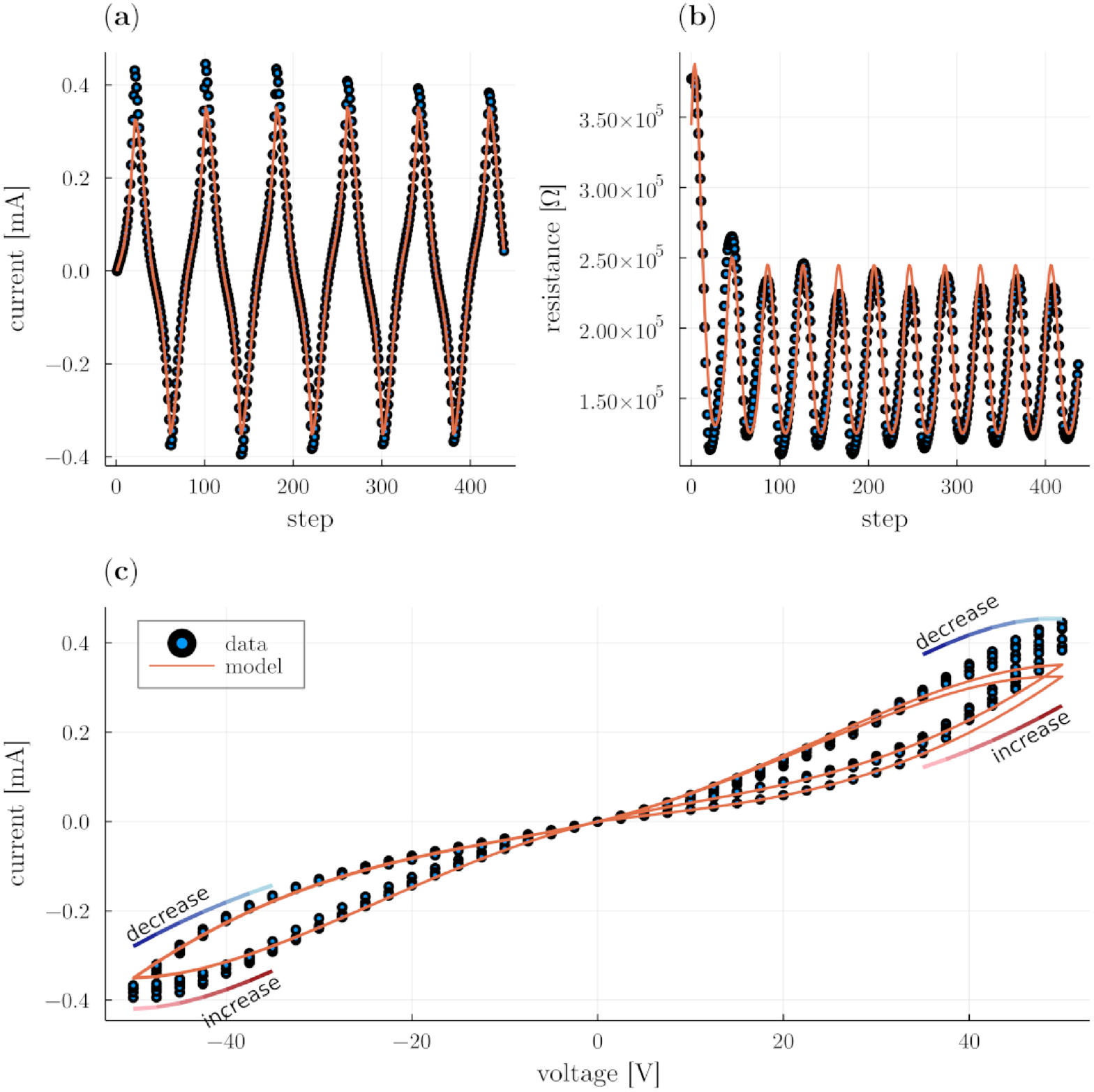}
\caption{\textit{ Phenomenological model. a) Current response, b) current-voltage characteristic and resistance changes for the measured data and the model fitted to $\tau = 0.93$, $\eta = 1.62e-8$ and $C_0 = 2.90e-6$ }}
\label{fig: fig_6}
\end{figure}

Based on the results discussed above, we propose a simplified discrete-time model of the system that displays the two key properties of our system: the continuous change of resistance and the direction of its changes.
For simplicity sake, the model is formalized as a difference-equation by considering time as discrete steps in which the current response $I_t$ is a function of conductance $C_t$ and voltage $V_t$.
The equations can be written as: 

\begin{equation}
\label{eq:3}
\begin{split}
 I_t &= C_t V_t \\
 C_t &= \tau C_{t-1} + \eta |V_t|
 \end{split}
\end{equation}

where $t$ corresponds to an iteration step and the coefficients $\eta$ and $\tau$ represent the voltage and time dependence of the change in conductance, respectively.
The parameters and the initial conductance value $C_0$ are $\tau = 0.93$, $\eta = 1.62e-8 mS/V$ and $C_0 = 2.90e-6 mS$ were estimated by fitting the model equations~\ref{eq:3} to the experimental data shown in Fig.~\ref{fig: fig_2}b).
The results presented in Fig.~\ref{fig: fig_6} show that the discrete model replicates well the key properties of the system: the continuous change in resistance and the direction of its changes.
Taking the absolute value in the voltage term $|$V$_t|$ is justified by the experimental data, because it exhibits symmetry regardless of the direction of the applied electric field.
The parameter $\eta$ is related to the absolute value of the voltage-dependent conductivity, which in this model is a constant.

On the contrary, in other models of memristive devices, based on resistance switching, the parameter $\eta$ may take three discrete values: 
\begin{equation}
\eta = \begin{cases}
0 &  V < V_\text{threshold}\\
c & V = V_\text{threshold}\\
-c & V = -V_\text{threshold}\\
\end{cases}
\end{equation}
\noindent where $c$ is a constant.
In terms of phenomenology, the physical interpretation of the model considers two components on the $C_t$ dynamics of Eq.~\eqref{eq:3} as two co-existing processes: the $\tau C_{t-1}$ component corresponding to relaxation processes and the $\eta|V_t|$ component related to the increase in conductance caused by the applied external electric field.
This view is consistent with our hypothesis, in which the external electric field forces ion migration/displacement and competes with spontaneous relaxation (i.e., volatility).
It is also worth noting that any other case than $\tau < 1$ would result in the disappearance of the hysteresis loop.
An extended analysis of the model will be discussed elsewhere.

\section{Conclusion}
We have developed a novel in-plane memristive device in which the memristive character originates indirectly from interactions between mobile ions in the PSS matrix and the dipole-induced electric field.
Processes that occur at the interface between PSS and P4VP cause dynamic changes in conductivity in R-P3HT.
We have shown, by step-by-step measurements, that the system does not manifest a memristive character in the absence of a layer with an ion reservoir, but we have also proven that there is no ionic conduction through the ion reservoir layer.
The memory effect in the system can be canceled by freezing the sample, which can be related to limiting the ion mobility at low temperatures.
The devices display smooth tuning of conductance caused by voltage sweeping or potentiation/depotentiation, which is exhibited in the form of highly stable hysteresis loops and short-term plasticity.
In addition, we have shown that to some extent we can control the dynamics of the system by changing the ionic composition of the reservoir. We have developed a discrete model  that replicates the unique shape and changes of conductivity in the hysteresis loop exhibited by our device.
These findings suggest that our system is a unique type of memristive device with, in contrast to resistive switching, dynamically altered conductance and with the ability to be manufactured in a cost-effective large-scale fabrication process, e.g., roll-to-roll processes, which makes the presented system an interesting research object.

% Experimental section

\section{Experimental Section}
\subsection{Materials}
All reagents were purchased from Sigma Aldrich, Alfa Aesar or Ossila and used without further purification or modification.
All materials were handled and processed in an inert atmosphere inside an argon-filled glovebox. \textbf{Caution need to be exercised since cobalt(II) bromide is toxic and should be handled with care}.

\subsection{Sample preparation}
All thin films were fabricated by spin casting of the respective material solutions.
Standard sensor substrates with ITO drain – source electrodes with channel dimensions of $30mm \times 50\mu m$ were used (Ossila S161).
Substrates were cleaned by sonication in 2-propanol for $15minute$ and dried under N$_2$ flux.
Electronic grade poly(3-hexylthiophene) (R-P3HT, Ossila M109 $95.2\%$RR, M$_w$ = 36,600, M$_n$ = 18,300, PDI = 2.0) was dissolved in chlorobenzene to form a $10mg/ml$ solution.
Before casting, the solution was heated at $55^{\circ}C$ overnight inside an argon-filled glovebox.
Before casting, the solution was filtered through a 0.45 $\mu$ m PTFE syringe filter.
Poly(4-vinylpyridine) was dissolved in pure ethanol (99.8$\%$).
The solution was prepared and casted inside an argon filled glovebox.
Ethanol was previously tested to not dissolve the R-P3HT layer.
Cobalt bromide was dissolved in acetonitrile to form 20 mg/ml solutions.
Acetonitrile was previously tested not to dissolve neither R-P3HT nor P4VP.
To introduce the metallic centers into the samples, each was immersed in the respective salt solution for $60s$.
The samples were then rinsed with pure acetonitrile to wash off the nonbound salt.
The $20 mg/ml$ PSS with addition of 1 mass $\%$ of potassium triflate was dissolved in deionized water (conductivity G = 0.07 $\mu$S), water was previously tested not to dissolve any element of the bilayer system.
It is crucial to ensure that there is no contact between the R-P3HT and PSS layers in order to achieve highly stable operation of the device.

\subsection{Electrical Characterization}
Direct current and current-voltage characteristics were measured under an argon atmosphere with a computer controlled Keithley 2400 source meter unit with a range of $-50$ to $50V$ with a $2.5V/s$ step.
Current-time measurements were performed under an argon atmosphere in voltage sequences: $-10;-50;+10;+50V$ with a duration of $20s$ of each potential step and a switching time below $25ms$.
Measurements for voltage sequences of $\pm10V$ and $\pm50V$ were performed with a duration of 6 s of each potential step and the switching time below $25ms$.
Temperature dependence measurements were performed in a vacuum chamber (pressure below 10$^{-6}$ mbar) with a Eurotherm 2416 temperature controller.
The sample was brought in contact with a cold finger to cool down to liquid nitrogen temperature; later, the contact was broken and the temperature of the sample slowly raised to room temperature without external heating.
Measurements were carried out at eight chosen points, at temperature ranges: 1) $-148.9$ to $-149.6^{\circ}C$, 2) $-123.5$ to $-122.5^{\circ}C$, 3) $-80.5$ to $-80.0^{\circ}C$, 4) $-54.0$ to $-52.0^{\circ}C$, 5) $-20.9$ to $-20.0^{\circ}C$, 6) $-2.5$ to $0^{\circ}C$, 7) $8.0$ to $9.0^{\circ}C$, and 8) $22.4$ to $22.5^{\circ}C$.

The current responses to the series of potentiating ( $+50V$) and depressing ( $-50V$) were measured in an argon atmosphere under $+10V$ bias with $100ms$ duration of each pulse and $225ms$ time interval between pulses.
The current response at various temperatures was measured inside a vacuum chamber using a computer-controlled Keithley 2400 source meter unit and Eurotherm 2416 temperature controller.

\subsection{Time–Of–Flight: Secondary Ion Mass Spectrometry}
The TOF-SIMS experiments were performed using TOF SIMS V an ION TOF GmbH instrument (Munster, Germany) equipped with a bismuth liquid metal ion gun used for surface analysis and a $2keV$ Cs ion source as a sputter gun.
Profiles, acquired in negative polarity, were collected in dual beam mode with Cs$^+$ $2keV$ as a sputtering beam together with Bi$_3^+$ as the analysis projectile configuration.
The angle of incidence of both ion beams was 45 degrees normal to the surface.
Each analysis was carried out in a $150 \times 150\mu m$ area concentric to the $300 \times 300 \mu m$ sputtering beam area.
\\
\subsection{Impedance Spectroscopy}
The impedance spectra PSS and PSS(+1\%K$^+$) were collected in a wide frequency range of $0.5$ to $3e6Hz$ with a measurement voltage of AC equal to 0.1 V by the Turnkey Impedance Spectrometer Concept 81 (Novocontrol Technologies GmbH $\&$ Co. KG, Montabaur, Germany).
All measurements were made at ambient temperature, and the architecture of the measured layers was the same as for the measurements of current-voltage characteristics.
Also, during impedance measurements, the layers were kept in an inert gas atmosphere (Ar).
As raw data, the impedance Z and phase angle $\phi$ were obtained directly, which were further recalculated on the real Z' and imaginary Z' parts of the impedance by WinDETA software.
To determine relaxation times related to sodium and potassium ions in PSS and PSS(+1\%K$^+$) layers, the data was fitted to the theoretical model (the Cole-Cole model) using Origin 2018 software, to both complex impedance parts in the frequency domain simultaneously.
Due to the ionic character of the dielectric relaxation observed on the dependence Z’, Z’’(f), the Cole-Cole model is a good enough approximation.
That ionic relaxation is not related to the relaxation of polymer chains, so, using the Havriliak-Negami model, it is not necessary.
Relaxation times were extracted from the relaxation frequency, which was obtained from the above-mentioned fitting procedure.

\bibliographystyle{unsrtnat}
\bibliography{article}  %%% Uncomment this line and comment out the ``thebibliography'' section below to use the external .bib file (using bibtex) .

\begin{thebibliography}{31}
\providecommand{\natexlab}[1]{#1}
\providecommand{\url}[1]{\texttt{#1}}
\expandafter\ifx\csname urlstyle\endcsname\relax
  \providecommand{\doi}[1]{doi: #1}\else
  \providecommand{\doi}{doi: \begingroup \urlstyle{rm}\Url}\fi

\bibitem[Jo et~al.(2010)Jo, Chang, Ebong, Bhadviya, Mazumder, and Lu]{Jo_2010}
Sung~Hyun Jo, Ting Chang, Idongesit Ebong, Bhavitavya~B. Bhadviya, Pinaki
  Mazumder, and Wei Lu.
\newblock Nanoscale memristor device as synapse in neuromorphic systems.
\newblock \emph{Nano Letters}, 10:\penalty0 1297--1301, 4 2010.
\newblock ISSN 1530-6984.
\newblock \doi{10.1021/nl904092h}.
\newblock URL \url{https://pubs.acs.org/doi/10.1021/nl904092h}.

\bibitem[Kim et~al.(2016)Kim, Sung, and Yoon]{Kim_2016}
Chang-Hyun Kim, Sujin Sung, and Myung-Han Yoon.
\newblock Synaptic organic transistors with a vacuum-deposited charge-trapping
  nanosheet.
\newblock \emph{Scientific Reports}, 6:\penalty0 33355, 12 2016.
\newblock ISSN 2045-2322.
\newblock \doi{10.1038/srep33355}.
\newblock URL \url{http://www.nature.com/articles/srep33355}.

\bibitem[Hodgkin and Huxley(1952)]{Hodgkin_1952}
A~L Hodgkin and A~F Huxley.
\newblock I952) i i7.
\newblock \emph{J. Physiol}, pages 500--544, 1952.

\bibitem[Chua(1971)]{Chua_1971}
L.~Chua.
\newblock Memristor-the missing circuit element.
\newblock \emph{IEEE Transactions on Circuit Theory}, 18:\penalty0 507--519,
  1971.
\newblock ISSN 0018-9324.
\newblock \doi{10.1109/TCT.1971.1083337}.
\newblock URL \url{http://ieeexplore.ieee.org/document/1083337/}.

\bibitem[Caravelli and Carbajal(2018)]{Caravelli_2018}
Francesco Caravelli and Juan~Pablo Carbajal.
\newblock Memristors for the curious outsiders.
\newblock \emph{Technologies}, 6\penalty0 (4), 2018.
\newblock ISSN 2227-7080.
\newblock \doi{10.3390/technologies6040118}.
\newblock URL \url{https://www.mdpi.com/2227-7080/6/4/118}.

\bibitem[Strukov et~al.(2008)Strukov, Snider, Stewart, and
  Williams]{Strukov_2008}
Dmitri~B. Strukov, Gregory~S. Snider, Duncan~R. Stewart, and R.~Stanley
  Williams.
\newblock The missing memristor found.
\newblock \emph{Nature}, 453:\penalty0 80--83, 5 2008.
\newblock ISSN 0028-0836.
\newblock \doi{10.1038/nature06932}.
\newblock URL \url{http://www.nature.com/articles/nature06932}.

\bibitem[Vongehr and Meng(2015)]{Vongehr_2015}
Sascha Vongehr and Xiangkang Meng.
\newblock The missing memristor has not been found.
\newblock \emph{Scientific Reports}, 5, 6 2015.
\newblock ISSN 20452322.
\newblock \doi{10.1038/srep11657}.

\bibitem[Pickett and Williams(2012)]{Pickett_2012}
Matthew~D Pickett and R~Stanley Williams.
\newblock Sub-100 fj and sub-nanosecond thermally driven threshold switching in
  niobium oxide crosspoint nanodevices.
\newblock \emph{Nanotechnology}, 23:\penalty0 215202, 6 2012.
\newblock ISSN 0957-4484.
\newblock \doi{10.1088/0957-4484/23/21/215202}.
\newblock URL
  \url{https://iopscience.iop.org/article/10.1088/0957-4484/23/21/215202}.

\bibitem[Jo et~al.(2009)Jo, Kim, and Lu]{Jo_2009}
Sung~Hyun Jo, Kuk~Hwan Kim, and Wei Lu.
\newblock Programmable resistance switching in nanoscale two-terminal devices.
\newblock \emph{Nano Letters}, 9:\penalty0 496--500, 2009.
\newblock ISSN 15306984.
\newblock \doi{10.1021/nl803669s}.

\bibitem[Xu et~al.(2019)Xu, Jang, Lee, Amanov, Cho, Kim, Park, Shin, and
  Ham]{Xu_2019}
Renjing Xu, Houk Jang, Min~Hyun Lee, Dovran Amanov, Yeonchoo Cho, Haeryong Kim,
  Seongjun Park, Hyeon~Jin Shin, and Donhee Ham.
\newblock Vertical mos2 double-layer memristor with electrochemical
  metallization as an atomic-scale synapse with switching thresholds
  approaching 100 mv.
\newblock \emph{Nano Letters}, 19:\penalty0 2411--2417, 2019.
\newblock ISSN 15306992.
\newblock \doi{10.1021/acs.nanolett.8b05140}.

\bibitem[Bessonov et~al.(2015)Bessonov, Kirikova, Petukhov, Allen, Ryhänen,
  and Bailey]{Bessonov_2015}
Alexander~A. Bessonov, Marina~N. Kirikova, Dmitrii~I. Petukhov, Mark Allen,
  Tapani Ryhänen, and Marc~J.A. Bailey.
\newblock Layered memristive and memcapacitive switches for printable
  electronics.
\newblock \emph{Nature Materials}, 14:\penalty0 199--204, 2015.
\newblock ISSN 14764660.
\newblock \doi{10.1038/nmat4135}.

\bibitem[Zeng et~al.(2014)Zeng, Li, Yang, Pan, and Guo]{Zeng_2014}
Fei Zeng, Sizhao Li, Jing Yang, Feng Pan, and D.~Guo.
\newblock Learning processes modulated by the interface effects in a
  ti/conducting polymer/ti resistive switching cell.
\newblock \emph{RSC Advances}, 4:\penalty0 14822, 2014.
\newblock ISSN 2046-2069.
\newblock \doi{10.1039/c3ra46679e}.
\newblock URL \url{http://xlink.rsc.org/?DOI=c3ra46679e}.

\bibitem[Goswami et~al.(2017)Goswami, Matula, Rath, Hedström, Saha, Annamalai,
  Sengupta, Patra, Ghosh, Jani, Sarkar, Motapothula, Nijhuis, Martin, Goswami,
  Batista, and Venkatesan]{Goswami_2017}
Sreetosh Goswami, Adam~J. Matula, Santi~P. Rath, Svante Hedström, Surajit
  Saha, Meenakshi Annamalai, Debabrata Sengupta, Abhijeet Patra, Siddhartha
  Ghosh, Hariom Jani, Soumya Sarkar, Mallikarjuna~Rao Motapothula, Christian~A.
  Nijhuis, Jens Martin, Sreebrata Goswami, Victor~S. Batista, and
  T.~Venkatesan.
\newblock Robust resistive memory devices using solution-processable
  metal-coordinated azo aromatics.
\newblock \emph{Nature Materials}, 16:\penalty0 1216--1224, 12 2017.
\newblock ISSN 1476-1122.
\newblock \doi{10.1038/nmat5009}.
\newblock URL \url{http://www.nature.com/articles/nmat5009}.

\bibitem[Xiao and Huang(2016)]{Xiao_2016}
Zhengguo Xiao and Jinsong Huang.
\newblock Energy‐efficient hybrid perovskite memristors and synaptic devices.
\newblock \emph{Advanced Electronic Materials}, 2:\penalty0 1600100, 7 2016.
\newblock ISSN 2199-160X.
\newblock \doi{10.1002/aelm.201600100}.
\newblock URL \url{https://onlinelibrary.wiley.com/doi/10.1002/aelm.201600100}.

\bibitem[Alibart et~al.(2010)Alibart, Pleutin, Guérin, Novembre, Lenfant,
  Lmimouni, Gamrat, and Vuillaume]{Alibart_2010}
Fabien Alibart, Stéphane Pleutin, David Guérin, Christophe Novembre,
  Stéphane Lenfant, Kamal Lmimouni, Christian Gamrat, and Dominique Vuillaume.
\newblock An organic nanoparticle transistor behaving as a biological spiking
  synapse.
\newblock \emph{Advanced Functional Materials}, 20:\penalty0 330--337, 1 2010.
\newblock ISSN 1616301X.
\newblock \doi{10.1002/adfm.200901335}.
\newblock URL \url{https://onlinelibrary.wiley.com/doi/10.1002/adfm.200901335}.

\bibitem[Younis et~al.(2013)Younis, Chu, Lin, Yi, Dang, and Li]{Younis_2013}
Adnan Younis, Dewei Chu, Xi~Lin, Jiabao Yi, Feng Dang, and Sean Li.
\newblock High-performance nanocomposite based memristor with controlled
  quantum dots as charge traps.
\newblock \emph{ACS Applied Materials \& Interfaces}, 5:\penalty0 2249--2254, 3
  2013.
\newblock ISSN 1944-8244.
\newblock \doi{10.1021/am400168m}.
\newblock URL \url{https://pubs.acs.org/doi/10.1021/am400168m}.

\bibitem[Liu et~al.(2016)Liu, Wang, Zhang, Pan, Zhang, Yang, Fan, Chen, and
  Li]{Liu_2016}
Gang Liu, Cheng Wang, Wenbin Zhang, Liang Pan, Chaochao Zhang, Xi~Yang, Fei
  Fan, Yu~Chen, and Run-Wei Li.
\newblock Organic biomimicking memristor for information storage and processing
  applications.
\newblock \emph{Advanced Electronic Materials}, 2:\penalty0 1500298, 2 2016.
\newblock ISSN 2199160X.
\newblock \doi{10.1002/aelm.201500298}.
\newblock URL \url{https://onlinelibrary.wiley.com/doi/10.1002/aelm.201500298}.

\bibitem[Park et~al.(2020)Park, Kim, Kim, and Lee]{Park_2020}
Hea-Lim Park, Min-Hwi Kim, Min-Hoi Kim, and Sin-Hyung Lee.
\newblock Reliable organic memristors for neuromorphic computing by predefining
  a localized ion-migration path in crosslinkable polymer.
\newblock \emph{Nanoscale}, 12:\penalty0 22502--22510, 11 2020.
\newblock ISSN 2040-3364.
\newblock \doi{10.1039/D0NR06964G}.
\newblock URL \url{http://xlink.rsc.org/?DOI=D0NR06964G}.

\bibitem[Zhu et~al.(2014)Zhu, Wan, Guo, Shi, and Wan]{Zhu_2014}
Li~Qiang Zhu, Chang~Jin Wan, Li~Qiang Guo, Yi~Shi, and Qing Wan.
\newblock Artificial synapse network on inorganic proton conductor for
  neuromorphic systems.
\newblock \emph{Nature Communications}, 5:\penalty0 3158, 5 2014.
\newblock ISSN 2041-1723.
\newblock \doi{10.1038/ncomms4158}.
\newblock URL \url{http://www.nature.com/articles/ncomms4158}.

\bibitem[Driscoll et~al.(2009)Driscoll, Kim, Chae, Ventra, and
  Basov]{Driscoll_2009}
T.~Driscoll, H.-T. Kim, B.-G. Chae, M.~Di Ventra, and D.~N. Basov.
\newblock Phase-transition driven memristive system.
\newblock \emph{Applied Physics Letters}, 95:\penalty0 043503, 7 2009.
\newblock ISSN 0003-6951.
\newblock \doi{10.1063/1.3187531}.
\newblock URL \url{http://aip.scitation.org/doi/10.1063/1.3187531}.

\bibitem[Tuma et~al.(2016)Tuma, Pantazi, Gallo, Sebastian, and
  Eleftheriou]{Tuma_2016}
Tomas Tuma, Angeliki Pantazi, Manuel~Le Gallo, Abu Sebastian, and Evangelos
  Eleftheriou.
\newblock Stochastic phase-change neurons.
\newblock \emph{Nature Nanotechnology}, 11:\penalty0 693--699, 8 2016.
\newblock ISSN 1748-3387.
\newblock \doi{10.1038/nnano.2016.70}.
\newblock URL \url{http://www.nature.com/articles/nnano.2016.70}.

\bibitem[Ge et~al.(2018)Ge, Wu, Kim, Shi, Sonde, Tao, Zhang, Lee, and
  Akinwande]{Ge_2018}
Ruijing Ge, Xiaohan Wu, Myungsoo Kim, Jianping Shi, Sushant Sonde, Li~Tao,
  Yanfeng Zhang, Jack~C. Lee, and Deji Akinwande.
\newblock Atomristor: Nonvolatile resistance switching in atomic sheets of
  transition metal dichalcogenides.
\newblock \emph{Nano Letters}, 18:\penalty0 434--441, 1 2018.
\newblock ISSN 1530-6984.
\newblock \doi{10.1021/acs.nanolett.7b04342}.
\newblock URL \url{https://pubs.acs.org/doi/10.1021/acs.nanolett.7b04342}.

\bibitem[Lu et~al.(2020)Lu, Sun, Liu, Li, Wang, Hao, Wang, Wang, and
  Zhang]{Lu_2020}
Qifeng Lu, Fuqin Sun, Lin Liu, Lianhui Li, Yingyi Wang, Mingming Hao, Zihao
  Wang, Shuqi Wang, and Ting Zhang.
\newblock Biological receptor-inspired flexible artificial synapse based on
  ionic dynamics.
\newblock \emph{Microsystems \& Nanoengineering}, 6:\penalty0 84, 12 2020.
\newblock ISSN 2055-7434.
\newblock \doi{10.1038/s41378-020-00189-z}.
\newblock URL \url{https://www.nature.com/articles/s41378-020-00189-z}.

\bibitem[Pickett et~al.(2013)Pickett, Medeiros-Ribeiro, and
  Williams]{Pickett_2013a}
Matthew~D. Pickett, Gilberto Medeiros-Ribeiro, and R.~Stanley Williams.
\newblock A scalable neuristor built with mott memristors.
\newblock \emph{Nature Materials}, 12:\penalty0 114--117, 2 2013.
\newblock ISSN 1476-1122.
\newblock \doi{10.1038/nmat3510}.
\newblock URL \url{http://www.nature.com/articles/nmat3510}.

\bibitem[Im et~al.(2020)Im, Kim, and Jang]{Im_2020}
In~Hyuk Im, Seung~Ju Kim, and Ho~Won Jang.
\newblock Memristive devices for new computing paradigms.
\newblock \emph{Advanced Intelligent Systems}, 2:\penalty0 2000105, 11 2020.
\newblock ISSN 2640-4567.
\newblock \doi{10.1002/aisy.202000105}.
\newblock URL \url{https://onlinelibrary.wiley.com/doi/10.1002/aisy.202000105}.

\bibitem[Wang et~al.(2017)Wang, Wang, Nagai, Xie, Yi, and Huang]{Wang_2017}
Zhiyong Wang, Laiyuan Wang, Masaru Nagai, Linghai Xie, Mingdong Yi, and Wei
  Huang.
\newblock Nanoionics‐enabled memristive devices: Strategies and materials for
  neuromorphic applications.
\newblock \emph{Advanced Electronic Materials}, 3:\penalty0 1600510, 7 2017.
\newblock ISSN 2199-160X.
\newblock \doi{10.1002/aelm.201600510}.
\newblock URL \url{https://onlinelibrary.wiley.com/doi/10.1002/aelm.201600510}.

\bibitem[Li et~al.(2013)Li, Zeng, Chen, Liu, Tang, Gao, Song, Lin, Pan, and
  Guo]{Li_2013}
Sizhao Li, Fei Zeng, Chao Chen, Hongyan Liu, Guangsheng Tang, Shuang Gao, Cheng
  Song, Yisong Lin, Feng Pan, and Dong Guo.
\newblock Synaptic plasticity and learning behaviours mimicked through ag
  interface movement in an ag/conducting polymer/ta memristive system.
\newblock \emph{Journal of Materials Chemistry C}, 1:\penalty0 5292, 9 2013.
\newblock ISSN 2050-7526.
\newblock \doi{10.1039/c3tc30575a}.
\newblock URL \url{http://xlink.rsc.org/?DOI=c3tc30575a}.

\bibitem[Mead(1990)]{Mead_1990}
Carver Mead.
\newblock Neuromorphic electronic systems.
\newblock \emph{Proceedings of the IEEE}, 78:\penalty0 1629--1636, 1990.
\newblock ISSN 1558-2256.
\newblock \doi{10.1109/5.58356}.

\bibitem[Majcher et~al.(2018)Majcher, Dąbczyński, Marzec, Ceglarska, Rysz,
  Bernasik, Ohkoshi, and Stefańczyk]{Majcher_2018}
Anna~M. Majcher, Paweł Dąbczyński, Mateusz~M. Marzec, Magdalena Ceglarska,
  Jakub Rysz, Andrzej Bernasik, Shin-ichi Ohkoshi, and Olaf Stefańczyk.
\newblock Between single ion magnets and macromolecules: a polymer/transition
  metal-based semi-solid solution.
\newblock \emph{Chem. Sci.}, 9:\penalty0 7277--7286, 2018.
\newblock \doi{10.1039/C8SC02277A}.
\newblock URL \url{http://dx.doi.org/10.1039/C8SC02277A}.

\bibitem[Dąbczyński et~al.(2020)Dąbczyński, Pawłowska, Majcher-Fitas,
  Stefańczyk, Dłubacz, Tomczyk, Marzec, Bernasik, Budkowski, and
  Rysz]{Dabczynski_2020}
Paweł Dąbczyński, Agnieszka~I. Pawłowska, Anna~M. Majcher-Fitas, Olaf
  Stefańczyk, Anna Dłubacz, Wojciech Tomczyk, Mateusz~M. Marzec, Andrzej
  Bernasik, Andrzej Budkowski, and Jakub Rysz.
\newblock Extraordinary conduction increase in model conjugated/insulating
  polymer system induced by surface located electric dipoles.
\newblock \emph{Applied Materials Today}, 21:\penalty0 100880, 12 2020.
\newblock ISSN 23529407.
\newblock \doi{10.1016/j.apmt.2020.100880}.
\newblock URL
  \url{https://linkinghub.elsevier.com/retrieve/pii/S2352940720303280}.

\bibitem[Habisreutinger et~al.(2018)Habisreutinger, Noel, and
  Snaith]{Habisreutinger_2018}
Severin~N. Habisreutinger, Nakita~K. Noel, and Henry~J. Snaith.
\newblock Hysteresis index: A figure without merit for quantifying hysteresis
  in perovskite solar cells.
\newblock \emph{ACS Energy Letters}, 3:\penalty0 2472--2476, 10 2018.
\newblock ISSN 23808195.
\newblock \doi{10.1021/acsenergylett.8b01627}.

\end{thebibliography}

\end{document}